\documentstyle[12pt]{article}
\input{epsf}
 
\title{Lattice QCD with 0, 2 and 4 Quark Flavors}

\author{Robert D. Mawhinney\thanks{The two flavor
calculation reported here was done at Columbia in collaboration with
Shailesh Chandrasekharan, Dong Chen, Norman H.~Christ, Weonjong Lee and
Decai Zhu.  The zero and four flavor calculations were done in
collaboration with Dong Chen and Norman Christ at Columbia and Gregory
W.~Kilcup at the Ohio State University.  This work was supported in
part by the U.S. Department of Energy.  Presented at RHIC Summer Study
'96, Brookhaven National Laboratory}
     \\
     Columbia University \\
     Department of Physics \\
     New York, NY~10027, USA}
\date{November 1, 1996}
\begin{document}
\maketitle

\begin{abstract}
\medskip                                                          
We have done simulations of lattice QCD with different numbers of light
dynamical quarks.  Since we cannot reach the continuum limit with our
current computers, we have done comparisons with 0 (quenched), 2 and 4
light quark flavors with the physical volume and lattice spacing
constant, when these are determined from the mass of the rho.  We find
a 7\% (2$\sigma$) difference in the nucleon to rho mass ratio for 2 and
4 quarks.  More importantly, the effects of chiral symmetry breaking
are dramatically decreased for the case of 4 light quarks.
\end{abstract}

\def\thepage{CU--TP--802}
\thispagestyle{myheadings}
\newpage
\pagenumbering{arabic}
\addtocounter{page}{1}

\section{Introduction}
\label{sec:introduction}

The role of light dynamical quarks in QCD is not fully understood.
Perturbatively, light quarks introduce screening and cause the QCD
coupling constant to evolve more slowly.  The role of light quarks in
the low-energy non-perturbative physics of QCD is much less certain.  Of
major interest in the numerical study of lattice QCD is a qualitative
and quantitative determination of light quark effects in low-energy
hadronic physics.

In lattice QCD, many calculations have been done in the quenched
approximation \cite{review}, where the effects of dynamical quarks are
removed.  Valence quarks can still be introduced, allowing hadron
masses to be measured from the decay of two-point hadronic correlation
functions, but there are no closed loops involving quarks.  Some
quenched calculations report hadron masses in reasonable agreement with
experiment but what is not known is whether this trend will continue to
weaker couplings.  In addition, theoretical arguments in the context of
chiral perturbation theory have given predictions for specific ways in
which the fermion truncation of the quenched approximation will appear;
quenched chiral logarithms.  If these quenched chiral logarithms appear
in numerical results they will certainly cloud the extraction of hadron
masses, etc.

Lattice QCD calculations with dynamical fermions are less advanced.
Inclusion of the fermionic determinant increases the required computer
power by at least 2 orders of magnitude.  Over the last few years,
calculations on $16^3 \times 32$, $16^3 \times 40$ and $20^4$ lattices
have been done by a number of groups \cite{review}.  However, a full
data set with various lattice spacings and volumes is not yet
available.  To date, very little difference has been seen between the
quenched and 2 flavor calculations at zero temperature for similar
volumes and lattice spacings.  (At finite temperature, differences
between the quenched and 2 flavor theories with staggered fermions have
been seen for a number of years.)

A study of full QCD comparable to the level of current quenched
calculations awaits the coming new computers which will push towards
the Teraflop scale.  Here we report a less ambitious calculation, which
still requires about 7 Gigaflop-years of computing;  a comparison of 0,
2 and 4 flavor QCD with a fixed lattice spacing and volume (in units of
the rho mass extrapolated to zero valence quark mass).  The
calculations we report only have results for a single dynamical fermion
mass, but we have calculated hadron masses and the chiral condensate
for a wide variety of valence quark masses.  As we detail below, even
from this restricted set of simulations, we have seen pronounced
effects of dynamical fermions for the four quark case.

Section \ref{sec:simulations} gives some details of our calculations.
and section \ref{sec:results} details the valence hadron masses we
measured for the three simulations and discusses the major differences
between them.  In section \ref{sec:finitev} we use a simple model of
finite volume effects to support our conclusions about the hadron mass
spectrum.

\section{Simulation Parameters}
\label{sec:simulations}

We are reporting on three different simulations whose parameters are
given in Table \ref{tab:parameters}.  The simulations were done
using the 16 Gigaflop, 256-node computer at Columbia, which is
just finishing its seventh year of full time computation.
(Additional quenched simulations were done by Greg Kilcup using
the T3D at the Ohio Supercomputer Center.  Some results from
these are reported in \cite{lat96}.)

\begin{table}
\caption{Parameters for the three simulations reported here.  The run
  length, thermalization, hadron measurement frequency and jackknife
  block size are in time units.}
\label{tab:parameters}
\begin{center}
\begin{tabular}{|l|c|c|c|} \hline
  			& $N_f=4$      & $N_f=2$      & $N_f=0$
			\\ \hline
  volume		& $16^3\times32$	& $16^3\times40$ 
			& $16^3\times32$	\\\hline
  $\beta$		& 5.4	& 5.7	& 6.05	\\\hline
  $m_{\rm dynamical} a$	& 0.01	& 0.01	& 	\\\hline
 	\hline
  evolution		& HMC	& HMD	& HMC	\\\hline
  run length		& 4450	& 4870	& 187,125 \\ \hline
  thermalization	& 250	& 250	& 375	\\ \hline
  acceptance rate 	& 0.95	&	& 0.91	\\\hline
  trajectory length	& 0.5	& 0.5	& 0.75	\\\hline
  step size		& 0.0078125	& 0.0078125 & 0.025 \\\hline
  CG stopping condition	& $1.13\times10^{-6}$	& $1.01\times10^{-5}$
			& 	\\\hline
   total run time	& 5 months	& 7.5 months
			& 1.7 months	\\\hline
\end{tabular}
\end{center}
\end{table}

All of our simulations were done with staggered fermions.  For 0 and 4
flavors, we used an exact hybrid Monte Carlo algorithm, while for 2
flavors, there is no practical exact algorithm available.  The R
algorithm we employed has ``$(\Delta t)^2$" errors, which means
observables will have systematic errors of this order, where for our
case $\Delta t = 0.0078125$.  The high acceptance rate for the exact
evolution, which has the same parameters as the inexact evolution
(except for the conjugate gradient stopping condition), demonstrates
that the $(\Delta t)^2$ errors are negligible for the inexact case.

We have measured hadron correlators using a variety of different source
sizes \cite{dch}, but only report results here from $16^3$ wall
sources.  With staggered fermions, the sinks which are used to select
the quantum numbers of the hadronic states can be, and in some cases
must be, non-local, due to the fact that the 4 components of the 4
fermionic flavors are delocalized.  We have used local sinks for all
the hadrons reported here.

\section{Results}
\label{sec:results}

\begin{figure}
  \vspace{-1in}
  \epsfxsize=\textwidth
  \epsffile{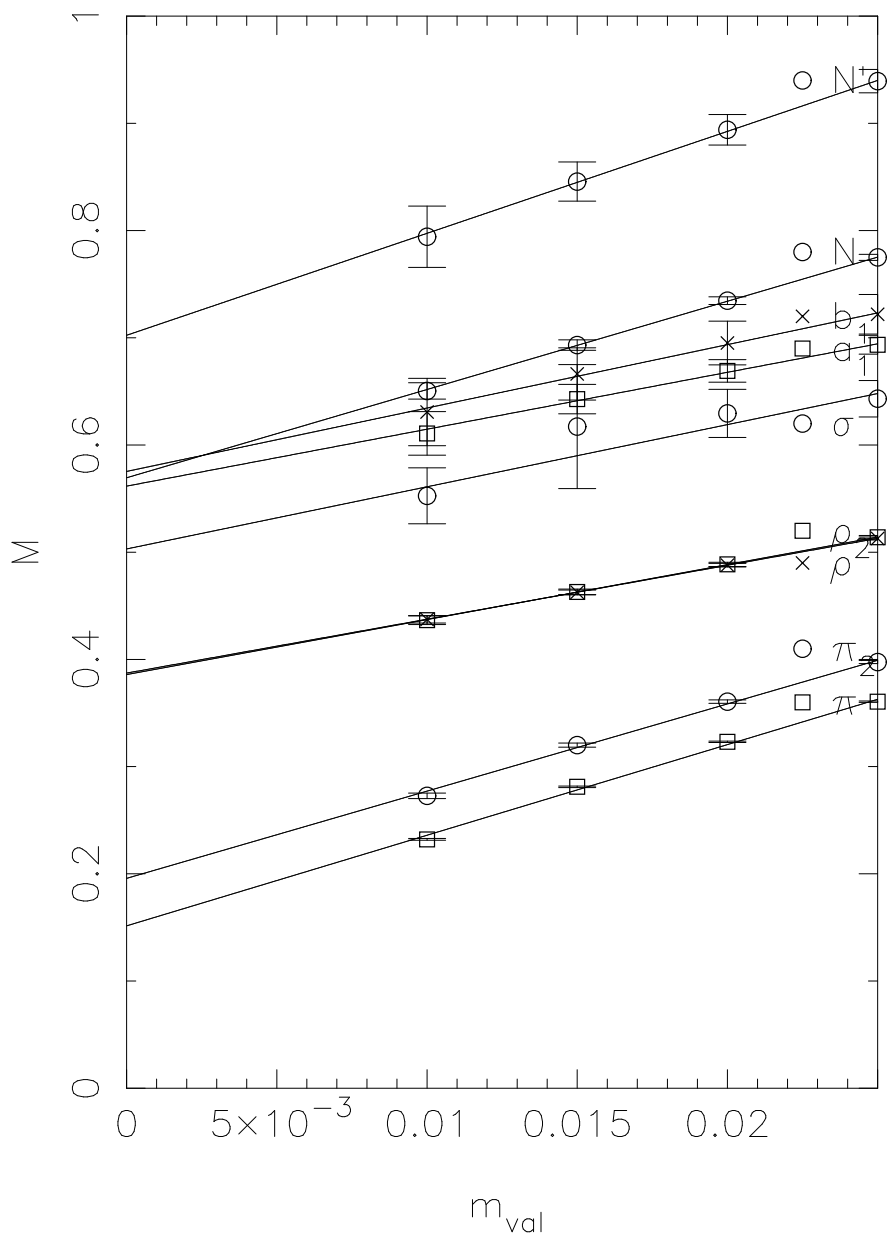}
  \caption{ Hadron masses versus $m_{\rm val}$ for the $16^3 \times 32 $
  quenched calculation at $\beta = 6.05$.}
  \label{fig:hm0}
\end{figure}

\begin{figure}
  \vspace{-1in}
  \epsfxsize=\textwidth
  \epsffile{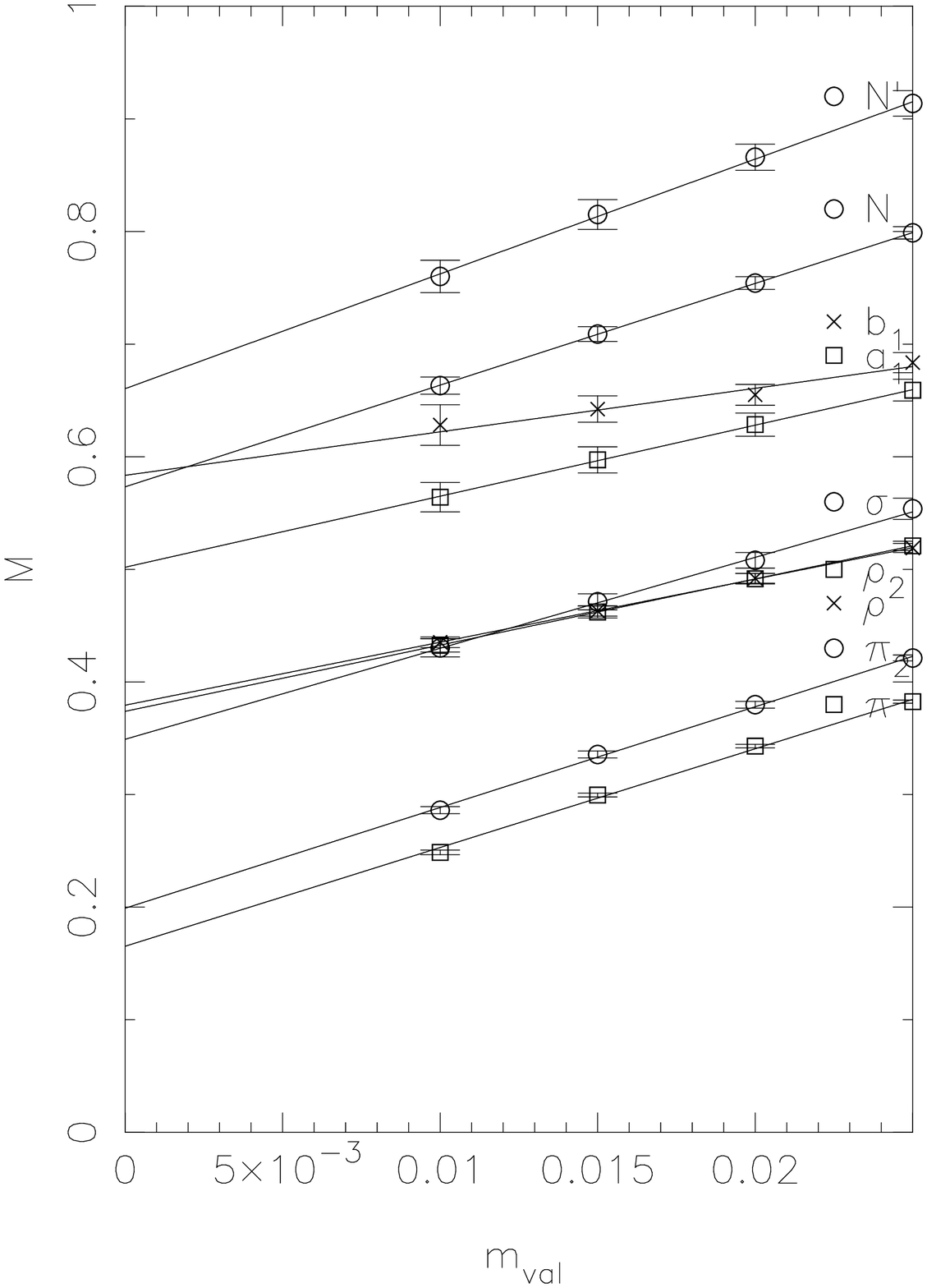}
  \caption{ Hadron masses versus $m_{\rm val}$ for the $16^3 \times 40 $
  two flavor calculation at $\beta = 5.7$ with $m_{\rm dyn} a = 0.01$.}
  \label{fig:hm2}
\end{figure}

\begin{figure}
  \vspace{-1in}
  \epsfxsize=\textwidth
  \epsffile{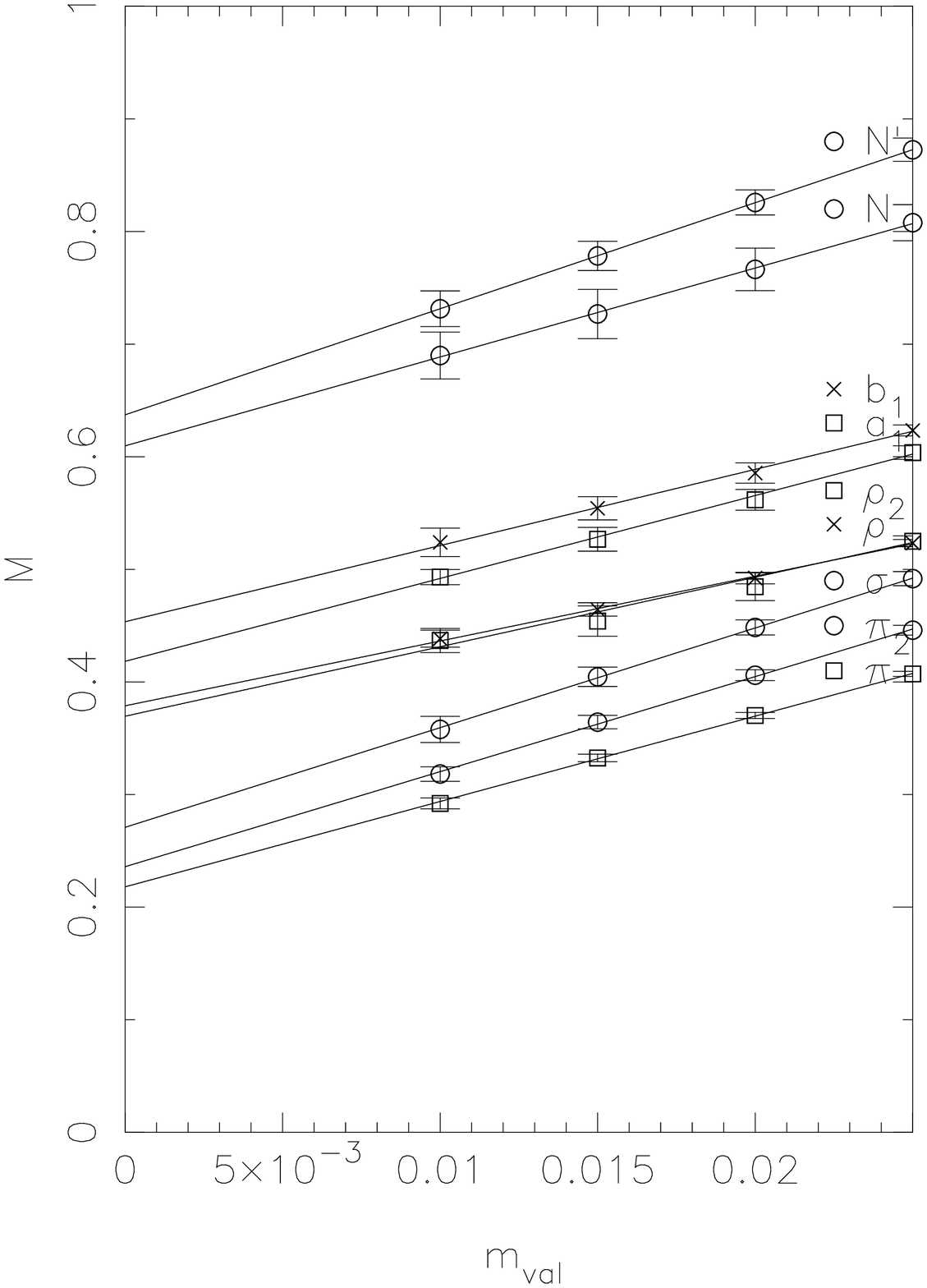}
  \caption{ Hadron masses versus $m_{\rm val}$ for the $16^3 \times 32 $
  four flavor calculation at $\beta = 5.4$ with $m_{\rm dyn} a = 0.01$.}
  \label{fig:hm4}
\end{figure}

Figures \ref{fig:hm0}, \ref{fig:hm2} and \ref{fig:hm4} are plots of
hadron masses as a function of the valence quark mass.  In order to
compare the plots, a few major points about staggered fermions should
be recalled.  At any finite lattice spacing, staggered fermions exhibit
flavor symmetry breaking.  The particles labeled $\rho$ and $\rho_2$
should become degenerate in the continuum limit and are actually quite
degenerate on the lattices we have studied.  The $a_1$ and $b_1$
correspond to the continuum mesons of the same names, as does the
nucleon $N$.  The $N'$ is the parity partner of the nucleon.


The $\pi$ and $\pi_2$ should also be degenerate in the limit of small
lattice spacing, but they are far from degenerate for the lattices we
studied.  An important feature of staggered fermions is the presence of
a U(1) chiral symmetry, for finite lattice spacing, when the quark mass
is set to zero.  (This U(1) symmetry is a subgroup of the SU(4)$_{\rm
A}$ flavor symmetry of four flavor continuum QCD and must not be
confused with the anomalous U(1)$_{\rm A}$ symmetry of the continuum.)
This U(1) chiral symmetry of staggered fermions then leads to a
Goldstone theorem for the valence pion and the prediction that
$m_{\pi}^2$ goes linearly to zero as the valence quark mass goes to
zero.  Note that it is $m_\pi$ versus $m_{\rm val}$ plotted here and
not $m_\pi^2$, which we will plot later.

The $\sigma$ particle in these figures is related to the staggered
fermion pseudo-Goldstone pion by a U(1) rotation on the valence quark
lines.  In the continuum, this $\sigma$ becomes a scalar, isoscalar
particle.  However, the $\sigma$ measured here does not include vacuum
bubble contributions.  Only quark propagators which start on the source
and end on the sink are included in the correlator.  This means that
for the case where $m_{\rm val}$ = $m_{\rm dyn}$, the $\sigma$ mass
reported here is not the mass for the scalar, isoscalar particle for
full QCD.  However, in the absence of chiral symmetry breaking, this
$\sigma$ and the $\pi$ are degenerate.

Comparing Figures \ref{fig:hm0}, \ref{fig:hm2} and \ref{fig:hm4}
reveals the following features:
\begin{enumerate}
\item  For all three calculations, the values of $\rho$ and $\rho_2$
agree quite closely for all valence masses and therefore in the
extrapolation to $m_{\rm val} = 0$.  The agreement at $m_{\rm val} = 0$
is a result of our choice of parameters;  we wanted to keep the physical
volume and lattice size constant in units of the $\rho$ mass.  The
fact that there is agreement for all valence masses was unexpected.
\item  While $m_\rho$ is the same for all three simulations, $m_N$
increases as more quarks are added.
\item  The mass of the $\sigma$ decreases as the number of quark
flavors is increased.  $m_\sigma$ is greater than $m_\rho$ for the
quenched calculation and clearly less than $m_\rho$ for four flavors.
\item  The splitting between parity partners ($\pi$, $\sigma$),
($\rho$, $a_1$) and ($N$, $N'$) decreases as the number of
dynamical quarks increases.  This is evidence for much smaller
chiral symmetry breaking effects as the number of
dynamical quarks is increased.  We will concentrate on this issue
in Section \ref{sec:finitev}
\end{enumerate}

\begin{figure}
  \vspace{-2in}
  \epsfxsize=\textwidth
  \epsffile{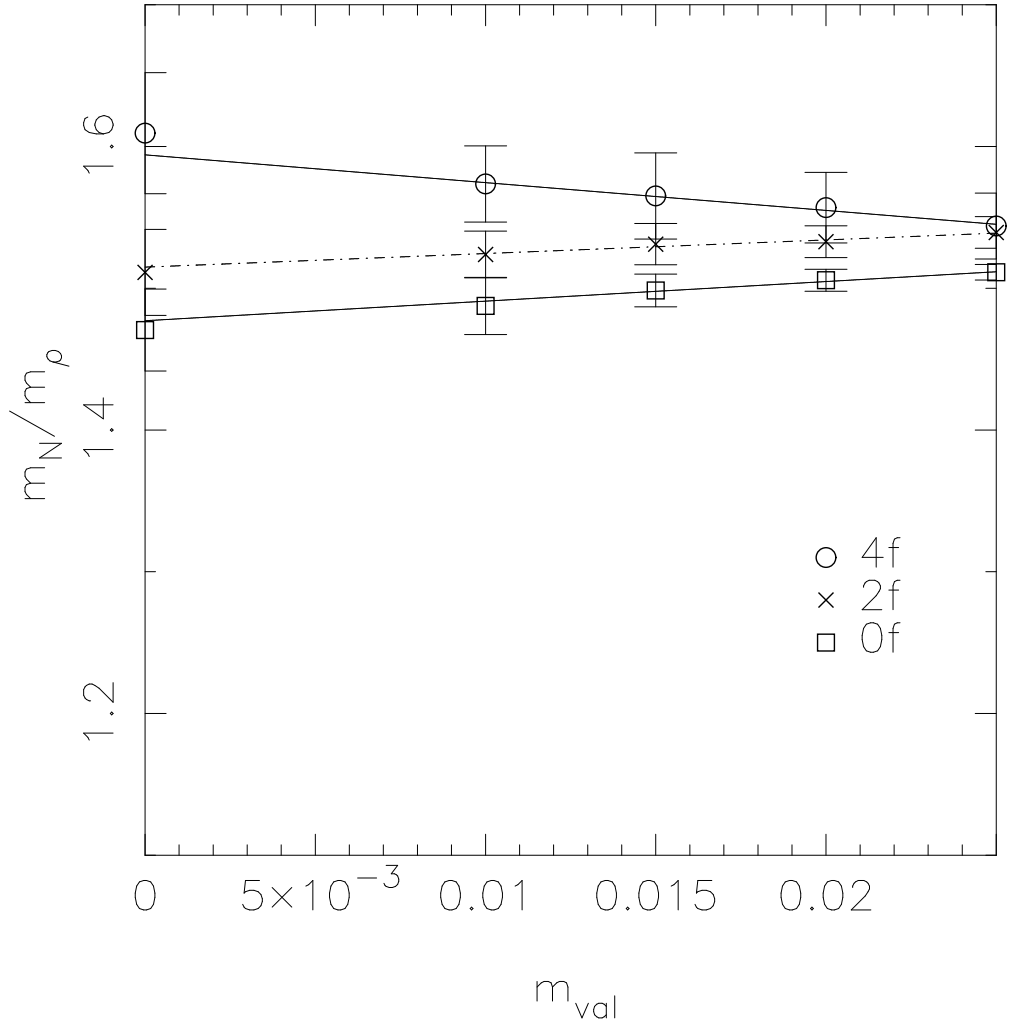}
  \caption{ $m_N / m_{\rho}$ vs.~$m_{\rm val}$ for the 0, 2 and 4 flavor
  calculations.  The points at $m_{\rm val} = 0 $ are
  the ratios of the extrapolated quantities, while the line is
  the extrapolation of the ratios.}
  \label{fig:nor}
\end{figure}

The variation in $m_N$/$m_\rho$ as the number of dynamical quarks
is increased is given in Figure \ref{fig:nor}.  There is very
little difference between the quenched and 2 flavor calculations
for this quantity.  This has been seen by others for $m_N$/$m_\rho$
and is also true of other zero-temperature hadronic quantities
like $B_K$.  However, the four flavor calculation is quite
distinct from the others.  The difference between the 2 and
4 flavor results is about 7\% or 2$\sigma$.  The statistical
difference of 2$\sigma$ does not make an ironclad case for a
difference, however as we have varied our statistical analysis
this is the smallest statistical difference we have seen.


\begin{figure}[t]
  \vspace*{-1.5in}
  \begin{center}
  \begin{minipage}[h]{5in}
    \epsfxsize=75mm
    \epsffile{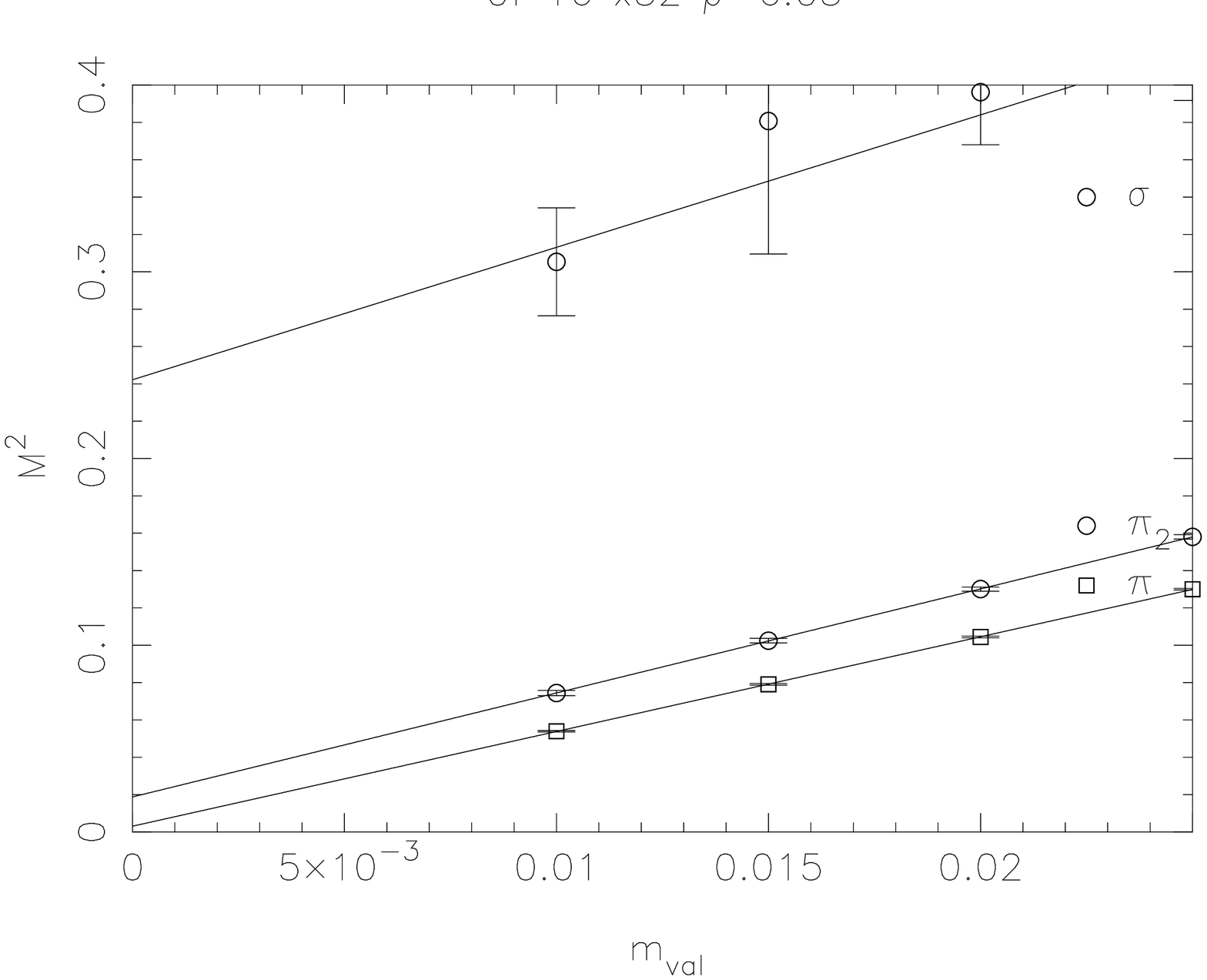}
  \end{minipage}
  \begin{minipage}[h]{3in}
    \vspace*{-1.3in}
    \epsfxsize=75mm
    \epsffile{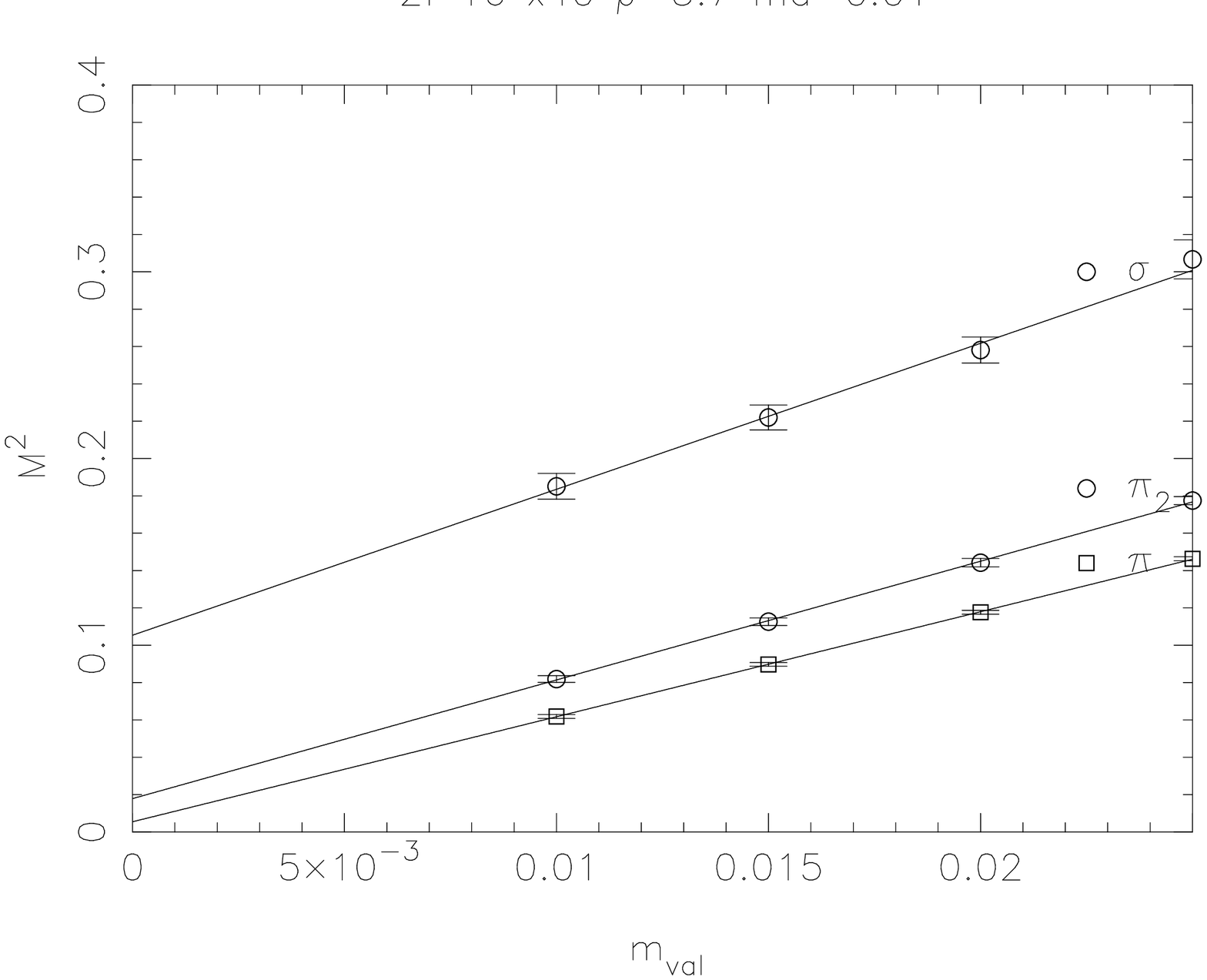}
  \end{minipage}
  \begin{minipage}[h]{3in}
    \vspace*{-1.3in}
    \epsfxsize=75mm
    \epsffile{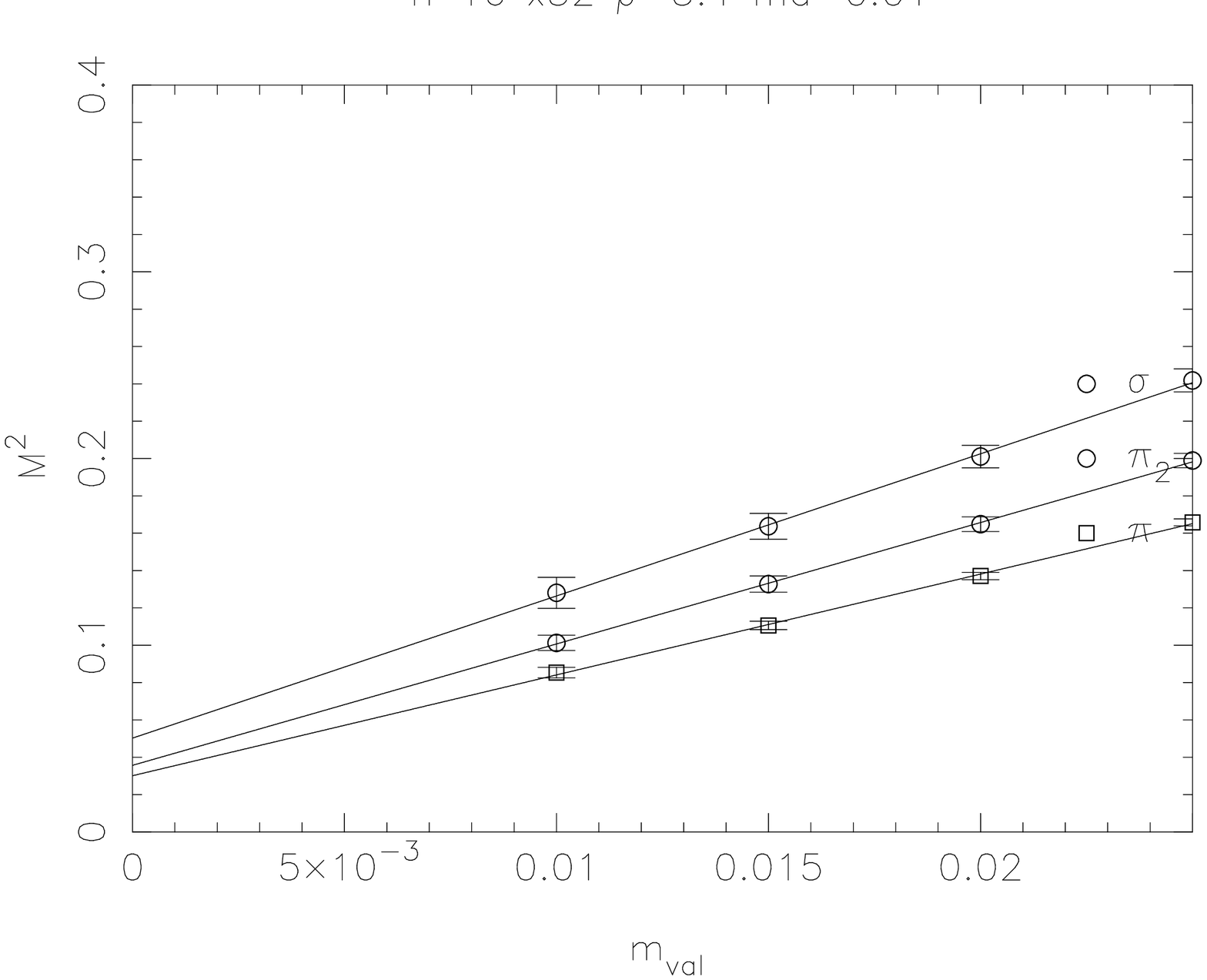}
  \end{minipage}
  \end{center}
  \vspace*{-1.5in}
  \caption{$m_{\pi}^2$, $m_{\pi_2}^2$ and $m_\sigma^2$
    vs.~$m_{\rm val}$ for the three simulations.}
  \label{fig:pi2}
\end{figure}

Another comparison between the three simulations is given in Figure
\ref{fig:pi2}, where $m^2_\pi$, $m_\sigma^2$ and $m^2_{\pi_2}$ are
plotted versus valence quark mass.  As mentioned earlier, the presence
of a Goldstone theorem for external quark lines means that $m_\pi^2$
should go to zero as $m_{\rm val} \rightarrow 0$.  This appears to be
the case for the quenched approximation, although the extrapolated value
for $m_\pi^2$ is actually 6 standard deviations away from zero.  For
the 2 and 4 flavor calculations, the intercept is growing relatively
larger, until for the 4 flavor calculation the intercepts for $m_\pi^2$
and $m_\sigma^2$ are closer to each other than the intercept for
$m_\pi^2$ is to zero.


To summarize, we have observed a systematic difference in
the value of $m_N/m_\rho$ between four flavors and the
other calculations.  The remaining features we observe in
the valence hadron spectrum are easily understood as a
decreased strength of chiral symmetry breaking, except the large
intercept for $m_\pi^2$ in the limit $m_{\rm val} \rightarrow 0$.
In the next section we use an earlier proposal of ours for finite
volume effects to understand this intercept.

\section{Chiral Symmetry Breaking and Finite Volume}
\label{sec:finitev}

Before drawing conclusions about the role of light dynamical fermions
in four flavor QCD, it is important to be confident that the four
flavor simulation we have done has the general qualitative properties
associated with the continuum limit of QCD, i.e. confinement and chiral
symmetry breaking.  There can be unexpected phase structure in theories
with many fermions \cite{nf8}, so we seek assurance that we are in a
phase consistent with continuum QCD. Evidence for confinement comes
indirectly through our hadron mass correlators;  we see propagating
states that fit the same functional forms as for two flavors where we
do have bound hadrons.  We have also measured Wilson lines
and find no evidence for deconfinement.

As mentioned above the four flavor hadron spectrum shows little effect
of chiral symmetry breaking.  In addition, extrapolations of
$m_\pi^2$ to $m_{val}=0$ give a large intercept, in apparent
contradiction to the Goldstone theorem.  This large intercept effect
has been known for some time in quenched simulations and was widely
expected to be caused by finite volume effects since the intercept
became closer to zero as the volume increased.  Recently
\cite{finitev}, we have proposed a simple model which can help to
quantify the role of finite volume effects in quenched or partially
quenched calculations. (Our 2 and 4 flavor calculations are sometimes
referred to as partially quenched calculations, since we are discussing
valence quark extrapolations on a set of gauge fields that were
generated including the effects of quark loops.)  To test whether the
four flavor results we are seeing are consistent with conventional QCD,
we now discuss the role of finite volume effects.

One simple effect of finite volume that has been predicted analytically
and is seen numerically is a cutoff in the average eigenvalue density
for the Dirac operator.  We can observe this effect numerically by
measuring the chiral condensate as a function of valence quark mass.
In a chirally asymmetric phase, when the valence quark mass is less
than the smallest eigenvalue of the Dirac operator, the valence chiral
condensate goes to zero linearly with the valence quark mass.

In particular, we write the valence chiral condensate
$\langle \bar{\zeta}\zeta \rangle$ as
\begin{equation}
 \langle \bar{\zeta}\zeta(m_{\zeta}) \rangle
 =
 2 m_{\zeta}
 \int_0^{\infty} \; d\lambda \; \frac{ \bar{\rho}(\lambda, \beta,
   m_{\rm dyn})} {\lambda^2 + m_{\zeta}^2} \label{eq:zcc_eigen}
\end{equation}
where $ \bar{\rho}(\lambda, \beta, m_{\rm dyn}) $ is the ensemble
average of the density of eigenvalues of the Dirac operator and
$m_{\zeta}$ is the valence mass for the quark fields $\zeta$ and
$\bar{\zeta}$.  (These fields do not enter in the dynamics; they are an
extra set of fermions used to probe the system.)  The ensemble average
can depend on the dynamical fermion mass used ($m_{\rm dyn}$) as well
as $\beta = 6/g^2$.  Our normalization is $\int \, d \lambda \,
\bar{\rho}(\lambda, \beta, m_{\rm dyn}) = 1$.  If $ \bar{\rho}(\lambda,
\beta, m_{\rm dyn}) $ is zero (or small) below some $\lambda_{\rm
min}$, then equation (\ref{eq:zcc_eigen}) gives $
 \langle \bar{\zeta}\zeta(m_{\zeta}) \rangle $ going linearly to zero.

To relate this finite volume effect in the quenched chiral condensate
to the intercept in the quenched pion mass squared. we use the fact
that there is a Gell-Mann--Oakes--Renner relation on the lattice, which
is independent of whether the ensemble of gauge fields is quenched
or unquenched,
\begin{equation}
  C(m_{\zeta})
  =
  \frac { m_{\pi}^2(m_{\zeta}) \; \langle \bar{\zeta}\zeta
    (m_{\zeta}) \rangle}
  { m_{\zeta}} \label{eq:gmo}
\end{equation}
where
\begin{equation}
  C(m_{\zeta}) = m_{\pi}^2(m_{\zeta}) \, \int \, d^4x \,
  \langle \pi(x) \pi(0) \rangle \label{eq:c_def}
\end{equation}
(The lattice version of this for staggered fermions is
$\langle \bar{\zeta}\zeta (m_{\zeta}) \rangle = m \sum_t
C_\pi(t)$  where $ C_\pi(t) $ is the staggered fermion pseudo-Goldstone
pion correlator \cite{toolkit}.)

\begin{figure}
  \vspace*{-0.5in}
  \begin{center}
  \begin{minipage}[t]{3in}
    \epsfxsize=75mm
    \epsffile{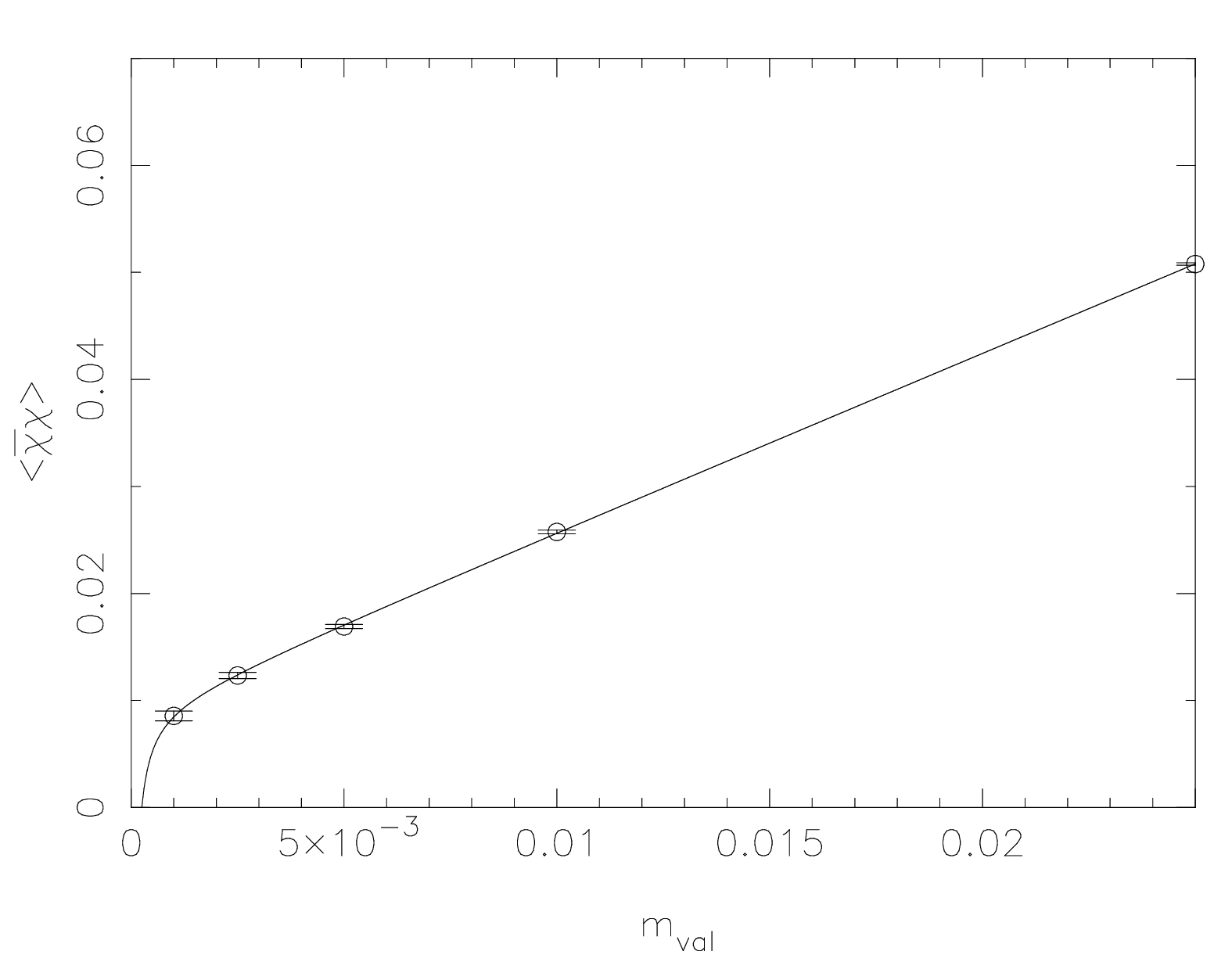}
  \end{minipage}
  \begin{minipage}[t]{3in}
    \epsfxsize=75mm
    \epsffile{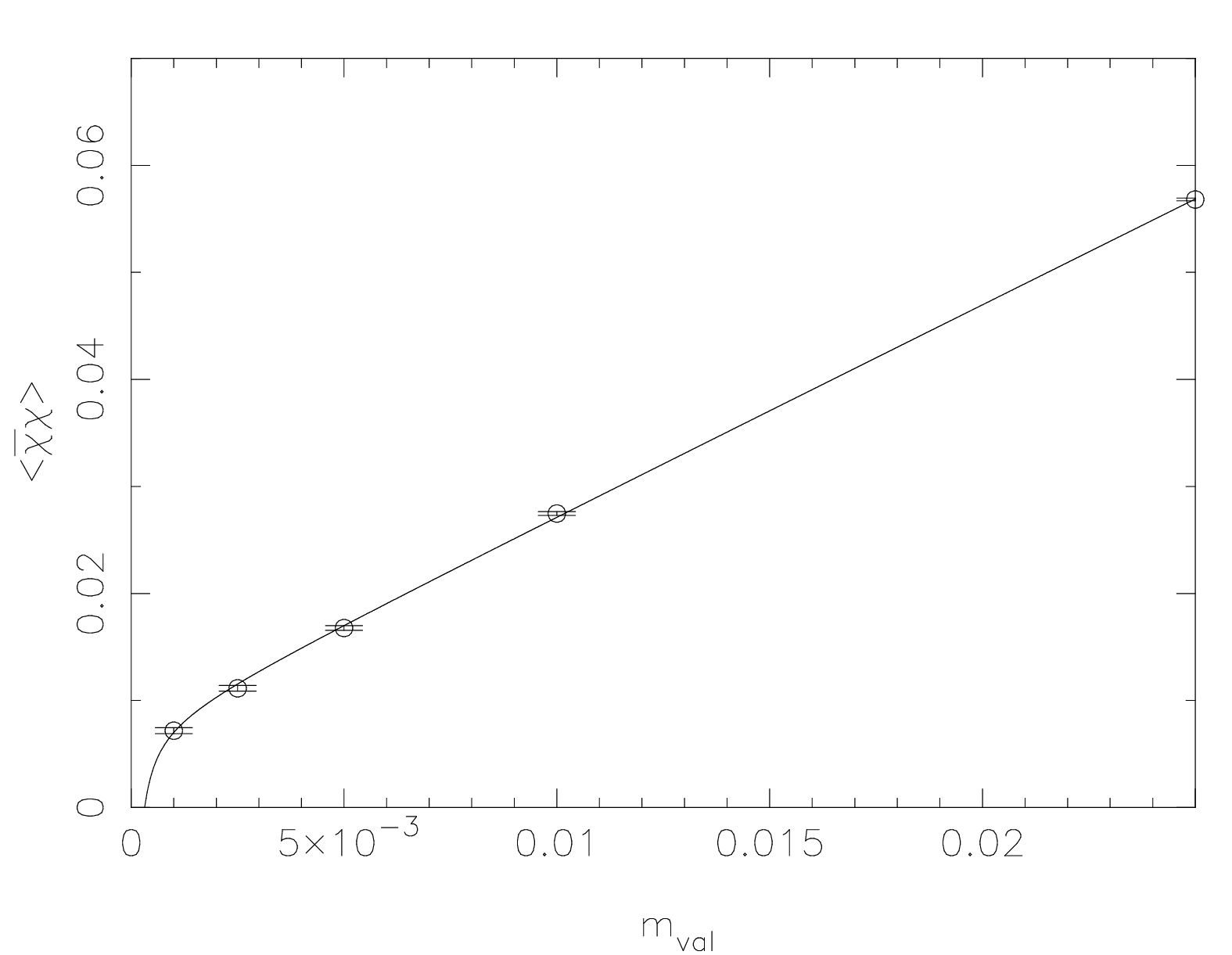}
  \end{minipage}
  \raisebox{-4ex}{}
  \begin{minipage}[h]{3in}
    \epsfxsize=75mm
    \epsffile{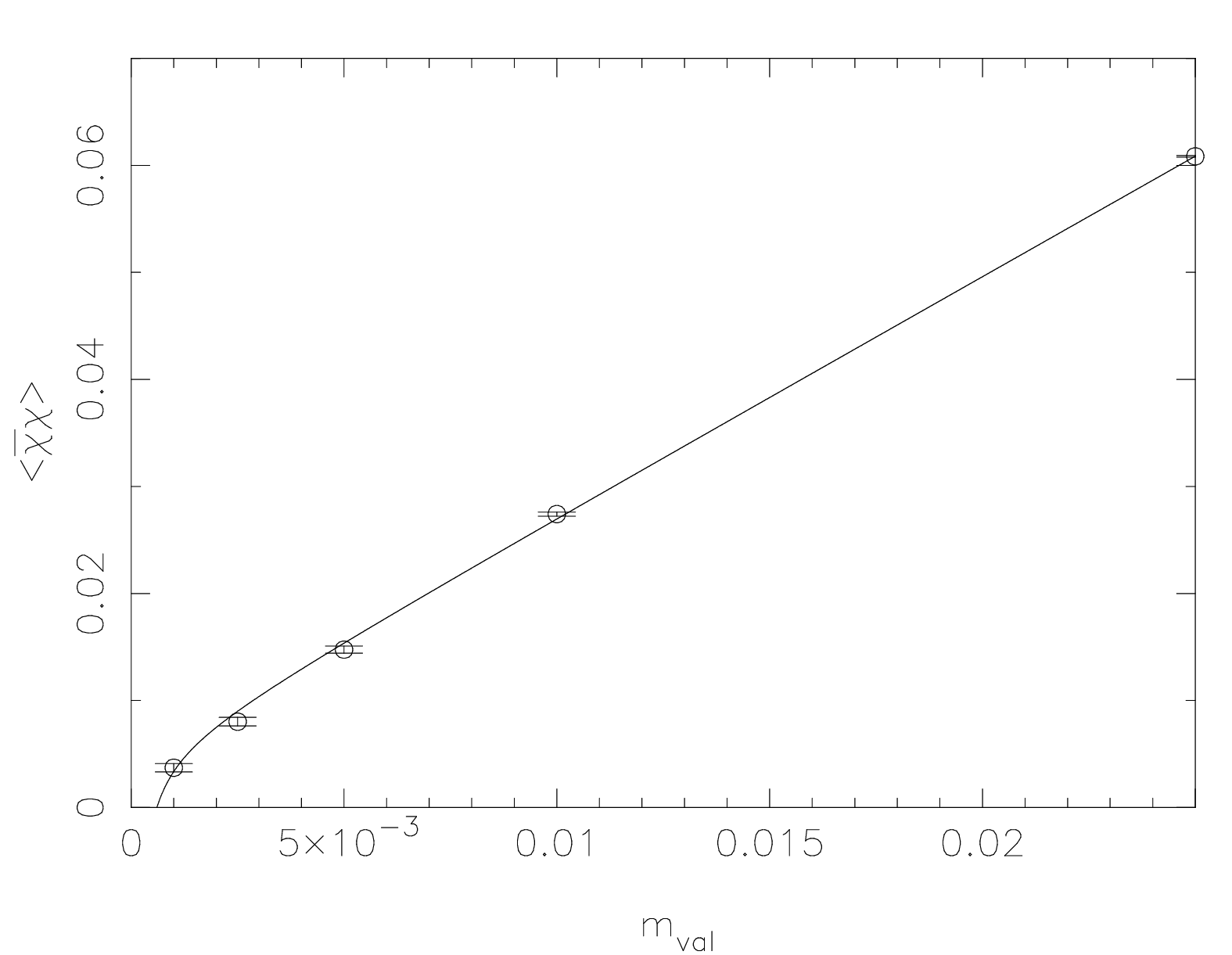}
  \end{minipage}
  \begin{minipage}[h]{3in}
  \end{minipage}
  \end{center}
  \caption{Fits of $\langle \bar{\zeta} \zeta \rangle$ to the form 
  $a_0 + a_{-1}/m_\zeta + a_1 m_\zeta$ for five values of
  $m_\zeta$.  The top figure is for 0 flavors, the middle for 2
  and the bottom for 4.}
  \label{fig:cc}
\end{figure}

As shown in \cite{finitev}, the assumption that $ \bar{\rho}(\lambda,
\beta, m_{\rm dyn}) $ drops dramatically below some $\lambda_{\rm min}$
and is constant for $\lambda_{\rm min} < \lambda < \lambda_0$ leads to
the result that
\begin{equation}
  \langle \bar{\zeta}\zeta(m_{\zeta}) \rangle = 
   a_0 + a_1 m_{\zeta} + a_{-1} / m_{\zeta} + O(m_\zeta^2)
   + 0(m_\zeta^{-2}), \label{eq:zcc_parm}
\end{equation}
for $ \lambda_{\rm min} < m_{\zeta} < \lambda_0$.  Figure \ref{fig:cc}
shows plots of  $ \langle \bar{\zeta}\zeta(m_{\zeta}) \rangle $
for our simulations.  The curvature at $m_\zeta \sim 10^{-4}$
is the onset of finite volume effects.


Now, assuming $C(m_{\zeta})$ is a smooth function of $m_\zeta$
in the range $ \lambda_{\rm min} < m_{\zeta} < \lambda_0$,
we have
\begin{equation}
  m_{\pi}^2 = - \frac{C(0) a_{-1}}{a_0^2} +
	\left[ \frac{C(0)}{a_0} - \frac{C^{\prime}(0)a_{-1}}
	{a_0^2} \right] m_{\zeta} + \cdots
  \label{eq:intercept}
\end{equation}
Thus a non-zero intercept for $m_\pi^2$ can be related to the finite
volume cutoff in the Dirac eigenvalue spectrum since $a_{-1} \sim
-\lambda_{\rm min}$.

Uncorrelated fits to our measurements of $\langle \bar{\zeta} \zeta
\rangle$ are given in Table \ref{tab:zcc_parm}.
\begin{table}
\caption{Fit parameters for fits to the form given in equation
  (4).}
\label{tab:zcc_parm}
\begin{center}
\begin{tabular}{|c|c|c|c|c|} \hline
  	& $a_0$	& $a_{-1}$ &  $a_1$ & $\chi^2$ \\ \hline

0f & 0.00917(23) & $-2.38(53) \times 10^{-6}$ & 1.669(10) & 1.0(17) 
 \\ \hline
2f & 0.00765(29) & $-2.61(17) \times 10^{-6}$ & 1.973(7) & 7.6(52)
 \\ \hline
4f & 0.00488(41) & $-3.77(19) \times 10^{-6}$ & 2.245(15) & 15(13)
 \\ \hline
\end{tabular}
\end{center}
\end{table}
The error on $\chi^2$ is the jackknifed error on this quantity.  For
four flavors, we find that $a_{-1}/a_0^2$ is about 3.6 times as large
as the two flavor value, while the ratio of the $m_\pi^2$ intercepts is
about 5.6.  There is clearly some $N_f$ dependence in $C(m_\zeta)$,
or higher order terms we have neglected in (\ref{eq:intercept}) are
important for good quantitative agreement.

This analysis supports the conclusion that we are in a phase with
chiral symmetry breaking, although the breaking is small.  The small
amount of breaking (small $a_0$) in a finite volume (non-zero $a_{-1}$)
leads to a large intercept for $m_\pi^2$.  We are investigating the
possibile $N_f$ dependence of $C(m_\zeta)$ and are working on a
determination of $f_\pi$ for the four flavor calculation.

\section{Conclusions}

We have seen that increasing the number of light dynamical quarks to
four does alter the valence hadron spectrum at zero temperature.  The
change in the nucleon to rho mass ratio is about 7\% and seems resolved
by our statistics.  A much larger effect is seen in the amount of
chiral symmetry breaking on the lattices.  The hadron spectrum exhibits
much less chiral symmetry breaking, which is consistent with the
suppression of small eigenvalues of the Dirac operator due to the
fermionic determinant.

By checking the role of finite volume effects in distorting
chiral symmetry breaking, we have argued that our four flavor
results are consistent with a chirally asymmetric, confining
theory.  To gain further evidence for the explanation proposed
here, calculations of the explicit eigenvalue density are
being done in collaboration with Robert Edwards from SCRI. 

Of great interest is whether the effects we have seen for
valence calculations done on a dynamical fermion background
persist when the dynamical and valence masses are varied
together.  We are currently undertaking another four flavor
calculation at a different dynamical quark mass to gain
some insight into this question.

{\bf Acknowledgements:} We would like to thank Catalin Malareanu for
measuring the Wilson lines on our lattices.


\begin{thebibliography}{99}

\bibitem{review} D. K. Sinclair, Nucl.\ Phys.\ {\bf B}
  (Proc.\ Suppl.) {\bf 47} (1996) 112; S. Gottlieb,
  hep-lat/9608107, to appear in the
  proceedings of Lattice '96.
\bibitem{lat96} D. Chen and R. D. Mawhinney, to appear in the
  proceedings of Lattice '96, St.\ Louis, June 1996.
\bibitem{dch} Dong Chen, Columbia University PhD Thesis,
  September, 1996.
\bibitem{nf8} F. R. Brown, {\it et.\ al.}, Phys.\ Rev.\ {\bf D46}
 (1992) 5655.
\bibitem{finitev} R. D. Mawhinney, Nucl.\ Phys.\ {\bf B}
  (Proc.\ Suppl.) {\bf 47} (1996) 557.
\bibitem{toolkit} G. W. Kilcup and S. R. Sharpe, Nucl.\ Phys.\
 {\bf B283} (1987) 493.
\end{thebibliography}
\end{document}